\documentclass[aps,prx,twocolumn,colorlinks=true,linkcolor=black,urlcolor=blue,citecolor=black,pdfhttps://www.overleaf.com/project/624ef482c726789d7c1a8f5cencoding=auto,longbibliography,superscriptaddress]{revtex4-2}
\usepackage{graphicx}
\usepackage{physics}
\usepackage{braket}
\usepackage{color}
\usepackage{amsmath}
\usepackage{amsfonts}
\usepackage{bbold}
\usepackage{amssymb}
\usepackage{tikz}
\usepackage{bm}
\usepackage{braket}
\usepackage{pgfplots}
\usepackage{dsfont}
\usepackage{comment}
\usepackage[colorlinks=true,
            citecolor=blue,
            pdfencoding=auto]{hyperref}
\usepackage{blindtext}
\usepackage{mathrsfs}
\usepackage[scr=boondoxo,scrscaled=1.05]{mathalfa}
\usepackage{lineno}
\usepackage{dirtytalk}

\begin{document}

\title{The Position Space Chern Number: A Topological Index for Chiral Magnetic Systems}

%Local Spectroscopy Around Trivializing Disorder Induced Chern Insulating Phase Transitions}

\author{Zachariah Addison}%
\email[Corresponding author: ]{za101@wellesley.edu}
\affiliation{Department of Physics and Astronomy,
Wellesley College,
Wellesley, MA 02482, USA}

\date{\today}% It is always \today, today,

\begin{abstract}
This paper introduces an index that categorizes the topology of insulating chiral magnetic systems.  The position space Chern number, $C_R$ is distinct from its momentum space counterpart, $C_K$.  A nonzero index guarantees the existence of topologically protected in-gap states that localize on the edge of local potential barriers in momentum space.  The Chern-Simons effective field theory describing position space Chern insulators reveals a topologically quantized correlation between transverse force operators that describe the flow of quanta in momentum space.  We demonstrate the existence of nonzero $C_R$ in systems hosting skyrmion magnetic phases and show how the index generalizes the classical concept of a skyrmion winding number.  Lastly we investigate the competition between momentum space and position space topologies, and highlight an apparent obstruction to having systems with both $C_R\neq0$ and $C_K\neq0$.    
\end{abstract}

\maketitle\

\section{Introduction}

Most topological insulators are theoretically described in the limit of a perfect atomic crystal.  Under periodic boundary conditions these systems possess a family of translation operators that commute with the Hamiltonian.  The eigenstates of these operators form a Bloch basis that can be indexed by a crystal momentum taking values in the first Brillouin zone \cite{bloch1929quantenmechanik}.  As such this basis serves as a natural space to investigate the topology of these systems \cite{qi2008topological,hasan2010colloquium,qi2011topological}.  However, in any real system these lattice translations are naturally broken by crystalline defects in the material or by emergent long wavelength potentials that derive from other interactions in the crystal.  In the presence of these perturbations it is less clear what would be the preferred basis to diagnose the topology of the system.

Here we highlight this ambiguity by investigating a {\it position space} Chern number, $C_R$, that is most naturally described in a local (or Wannier-like) basis \cite{wannier1937structure}.  Like the momentum space Chern number, $C_K$, this topological invariant is well defined for non-degenerate systems, is integer valued, can be associated with a bulk-edge correspondence, and results in a Chern-Simons  term in the effective action at long wavelengths \cite{haldane1988model,zee2007quantum,qi2008topological}.  This paper demonstrates these properties on model systems that naturally host $C_R\neq 0$ phases, and investigates the competition that can occur between momentum and position space topologies. 

The simplest systems hosting $C_R\neq 0$ are 2D or low dimensional chiral magnetic systems with a skyrmion magnetic phase \cite{roessler2006spontaneous, neubauer2009topological, nagaosa2013topological, fert2017magnetic, ahmed2019spin, shao2019topological,tokura2020magnetic}.  Skyrmion phases are described by a local magnetic texture, $\bm{m}(\bm{r})$ that winds the 2-sphere, $S^2$.  The winding (skyrmion) number (charge) associated with these textures, $N_\text{sk}$ can be calculated from $\bm{m}(\bm{r})$ via

\begin{equation}
    N_{\text{sk}}=\dfrac{1}{4\pi}\int d^2r\, \hat{\bm{m}}(\bm{r})\cdot \bigg(\pdv{\bm{\hat{m}}(\bm{r})}{x}(\bm{r})\times \pdv{\bm{\hat{m}}(\bm{r})}{y}\bigg)
    \label{NSK}
\end{equation}

\noindent
where $\hat{\bm{m}}(\bm{r})=\bm{m}(\bm{r})/|\bm{m}(\bm{r})|$ \cite{skyrme1962unified,belavin1975metastable,heinze2011spontaneous}.  For systems on periodic boundary conditions, equation \eqref{NSK} is integer valued, describing the solid angle swept out by $\hat{\bm{m}}(\bm{r})$ as it is evaluated across the system.  Skyrmion crystals appear naturally in the presence of mirror symmetry breaking that can allow for large Dzyaloshinskii–Moriya interactions \cite{nagaosa2013topological,wiesendanger2016nanoscale,tokura2020magnetic,back20202020}.

In the context of metals, skyrmion crystals are observed to have a topological Hall effect \cite{lee2009unusual,neubauer2009topological,kanazawa2011large,nagaosa2013topological,li2013robust,gallagher2017robust,ahmed2018chiral,ahmed2019spin,shao2019topological} that can be shown, theoretically, to couple directly to $N_{\text{sk}}$ in regimes where the skyrmion size $L_s\gg$ the lattice constant, $a$ \cite{verma2022unified,addison2023theory,addison2025anomalous}.  The manner in which $N_{\text{sk}}$ couples into the dynamics of conduction electrons in these materials is through a {\it semiclassical} position space Berry curvature, $\bm{\Omega}^{R,\text{sc}}(\bm{r},\bm{k})$ that leads to an anomalous force $\bm{F}\sim \bm{v}\times \bm{\Omega}^{R,\text{sc}}$, where $\bm{v}(\bm{k},\bm{r})$ is the semiclassical group velocity of a Bloch wavepacket centered at $\bm{r}$ and $\bm{k}$ \cite{xiao2010berry}.  For systems with periodic boundary conditions the integral of $\bm{\Omega}^{R,\text{sc}}$ across the system is a semiclassical position space Chern number, $C_{R}^{\text{sc}}$, that for weakly spin-orbit coupled systems is directly proportional to $N_{\text{sk}}$ \cite{addison2025anomalous}.

In this paper, we investigate the fully quantum versions of the position space Berry curvature and Chern number that can be associated with these semiclassical quantities.  In doing so, we present insulating systems whose topology can be indexed by $C_R$, and demonstrate the associated bulk-boundary correspondence for this topological index.  We also explore the effective Chern-Simons action for these systems and describe the physical consequences associated to these topological insulators.

\section{Position Space Chern Number}
\label{PSCN}

In this section we first review the momentum space Chern number and its relationship to the magnetic vector potential, and the global $U(1)$ gauge invariance of the wavefunction \cite{weyl1929electron}.  We then extend these ideas to define the analogous position space Chern number and determine some of its properties and relationships to quantities conjugate to variables associated to the momentum space Chern number.

The momentum space Chern number, $C_K$, naturally arises in condensed matter for nondegenerate two dimensional systems as the topological coefficient describing the size of the Hall conductivity, $\sigma_H=e^2C_K/h$ \cite{laughlin1981quantized,thouless1982quantized,niu1985quantized}.  The Hall conductivity describes the zero frequency homogeneous charge current $\bm{j}^Q$ that moves transverse to an applied constant electric field $\bm{E}$, $j_i^Q=\sum_{j=x,y}\sigma_{ij}E_j$ with $\sigma_H=(\sigma_{xy}-\sigma_{yx})/2$.  A constant electric field can be describe by a magnetic vector potential, $\bm{A}(t)$, that scales linearly in time ($\bm{E}(t)=-\partial_t\bm{A}(t))$.  As such, calculating, $\sigma_H$, amounts to understanding how $\bm{A}(t)$ perturbs the Hamiltonian, $\widehat{H}$, and the states of the system.

The electronic field operators have a global $U(1)$ gauge symmetry, $\widehat{\psi}(\bm{r},t)\rightarrow \alpha \widehat{\psi}(\bm{r},t)$, $\alpha\in U(1)$, that leads to a relationship between a conserved charge current, the magnetic vector potential, and the Hamiltonian: $\widehat{\bm{j}}^Q(\bm{r},t)=-\delta_r \widehat{H}/\delta_r \bm{A}(\bm{r},t)$, with the functional derivative defined as

\begin{equation}
    \dfrac{\delta}{\delta \bm{g}(\bm{r})}\bigg(\int d^2r' \, \bm{f}(\bm{r}')\cdot \bm{g}(\bm{r}')\bigg)=\bm{f}(\bm{r})
\end{equation}

\noindent
for some vector functions $\bm{f}(\bm{r})$ and $\bm{g}(\bm{r})$.  This can be understood as follows.  If the system has a global $U(1)$ gauge symmetry, then the actions $S$ must be left invariant under such at gauge transformation.  If instead we take $\alpha\rightarrow \alpha(\bm{r},t)$ then to linear order in $\alpha(\bm{r},t)$

\begin{align}
    \Delta S &\sim \sum_{\mu=0}^3\int d^3rdt \, \partial_{\mu}\alpha(\bm{r},t)h^\mu(\bm{r},t) \nonumber\\ &=-\sum_{\mu=0}^3\int d^3rdt \, \alpha(\bm{r},t)\partial_\mu h^\mu(\bm{r},t)
\end{align}

\noindent
where $\Delta S$ is the change in the action under the gauge transformation, $\partial_\mu=(\partial_t/c,\bm{\nabla}_r)$, and we have neglected boundary terms in line two.  For $\alpha$ constant, the gauge invariance demands that $\Delta S=0$ implying that $h^\mu$ satisfies a continuity equation $\sum_\mu \partial_\mu h^\mu=0$ and the existence of an underlying conserved quantity.

In the context of electronic charge, $\bm{A}(\bm{r},t)$ serves as a $U(1)$ gauge connection, hence the above relationship between the charge current, Hamiltonian, and magnetic vector potential.  Associated with the global gauge symmetry is the local conservation equation

\begin{equation}
    \pdv{n_r(\bm{r},t)}{t}+\bm{\nabla}_r\cdot \bm{j}^r(\bm{r},t)=0
    \label{conR}
\end{equation}

\noindent
where $n_r(\bm{r},t)=-\text{Tr}(\widehat{\rho}(t)\delta_r \widehat{H}/\delta_r e\phi(\bm{r}, t))=\bra{\bm{r}}\widehat{\rho}(t)\ket{\bm{r}}$ is the local number density and $\bm{j}^r(\bm{r},t)=-\text{Tr}(\widehat{\rho}(t)\delta_r \widehat{H}/\delta_r \hbar\bm{\mathcal{A}}^R(\bm{r},t))$ is the local number current with $\widehat{\rho}(t)$ the density matrix of the system, $\phi(\bm{r},t)$ the electromagnetic scalar potential, and $\bm{\mathcal{A}}^R(\bm{r},t)=e\bm{A}(\bm{r},t)/\hbar$.

Local gauge invariance of the system with respect to $\bm{A}(\bm{r},t) \rightarrow \bm{A}(\bm{r},t) +\bm{\nabla}_rf(\bm{r},t)$, for some function $\bm{f}(\bm{r},t)$, stipulates that the system couple to $\bm{A}(\bm{r},t)$ through the minimal coupling procedure of $\widehat{\bm{p}}\rightarrow \widehat{\bm{p}}+\hbar\bm{\mathcal{A}}^R(\widehat{\bm{r}},t)$ \cite{landau1975classical}.

The Hall conductivity is calculated by finding the expectation value of $\bm{\widehat{j}}^Q(t)$ to first order in $\bm{A}(t)$, which can be described by a current-current correlation function, by which then the momentum space Chern number can be determined ($C_K=h\sigma_H/e^2$).  In the thermodynamic limit at zero temperature ($T=0$), the associated momentum space Chern number can be written as

\begin{widetext}
\begin{equation}
    C_K=\dfrac{2\pi}{V}\sum_{nm} \dfrac{f_n-f_m}{(E_n-E_m)^2}\dfrac{i}{2}\bigg(\bra{n}\bm{\nabla}_{\mathcal{A}^R}\widehat{H}\ket{m}\times \bra{m}\bm{\nabla}_{\mathcal{A}^R}\widehat{H}\ket{n}\bigg) \cdot \bm{\hat{z}}
    \label{CKG}
\end{equation}
\end{widetext}

\noindent
where $\ket{n}$ are eigenstates of $\widehat{H}$ with eigenvalues $E_n$, $f_n=\theta(\mu-E_n)$ are the Fermi occupation functions for state $n$, and $V$ is the volume of the system \cite{kubo1957statistical,ando1975theory,streda1982theory,niu1985quantized}.  The derivatives $\bm{\nabla}_{\mathcal{A}^R}\widehat{H}$ are calculated by coupling the system to a homogeneous $\bm{\mathcal{A}}^R$, then taking $\bm{\mathcal{A}}^R\rightarrow 0$ after computing $\bm{\nabla}_{\mathcal{A}^R}\widehat{H}$.  See section \ref{CRCalc} for details of this calculation for systems on a lattice.  Equation \eqref{CKG} is well defined for systems with a non-degenerate spectrum for which $E_n\neq E_m$ for all $n,m$ with $f_n\neq f_m$.  In the thermodynamic limit at zero temperature, $C_K$ is integer valued and insensitive to perturbations that keep states from crossing the chemical potential $\mu$ \cite{niu1985quantized}.

Equation \eqref{CKG} has been used to calculate $C_K$, particularly for systems that lack translation symmetries, whether from disorder or the inclusion of electron-electron interactions, for which the Bloch Hamiltonian is ill-defined \cite{bellissard1994noncommutative, prodan2010entanglement,prodan2011disordered,hastings2011topological,bianco2011mapping,marrazzo2017locality,bourne2018non,caio2019topological,varjas2020computation,addison2025local}.

In an analogous fashion, we may identify a conserved current in momentum space by taking $\alpha\rightarrow \alpha(\bm{p},t)$. The corresponding field of interest we call $\bm{\mathcal{A}}^K(\bm{p},t)$, such that there is a conserved current $\bm{j}^p(\bm{p},t)$ satisfying

\begin{equation}
    \pdv{n_p(\bm{p},t)}{t}+\bm{\nabla}_p\cdot \bm{j}^p(\bm{p},t)=0
    \label{conP}
\end{equation}

\noindent
where $n_p(\bm{p},t)=\bra{\bm{p}}\widehat{\rho}(t)\ket{\bm{p}}$ is the number of particles per unit momentum volume, and $\bm{j}^p(\bm{p},t)=-\Tr(\widehat{\rho}(t)\delta_p \widehat{H}/\delta_p \bm{\mathcal{A}}^K(\bm{p},t))$, where $\bm{\mathcal{A}}^K(\bm{p},t)$ couples to $\widehat{H}$ via $\widehat{\bm{r}}\rightarrow \widehat{\bm{r}}+\bm{\mathcal{A}}^K(\widehat{\bm{p}},t)$.  Alternative definitions for $\bm{j}^p(\bm{p},t)$ have been studied perviously \cite{skodje1989flux,steuernagel2013wigner,nalewajski2015probability,nalewajski2015quantum,kakofengitis2017wigner,valtierra2020quasiprobability}, but do not satisfy a sourceless continuity equation and are not related to the global gauge invariance introduced above (see appendix \ref{JPA} for details).

The continuity equation can be understood as describing the relationship between the change in time of the number of particles having some momenta $\bm{p}$ in some volume of momentum space $V_p$, and the number current $\bm{j}^p(\bm{p},t)$ across the boundary of $V_p$.  In order to cross this boundary the momentum of the particle must change, indicating an applied force.  Note the units of $\bm{j}^p(\bm{p})$ are that of a force density.  In this sense $\bm{j}^p(\bm{p},t)$ and $\bm{j}^r(\bm{p},t)$ are conjugate quantities: $\bm{j}^r\sim \pdv{\bm{r}}{t}$ and $\bm{j}^p\sim \pdv{\bm{p}}{t}$.  In addition $n_r(\bm{r},t)$ and $n_p(\bm{p},t)$ are related to each other via

\begin{equation}
    n_p(\bm{p},t)= \dfrac{1}{(2\pi\hbar)^2}\iint d^2r_1 d^2r_2 \bra{\bm{r}_1}\widehat{\rho}(t)\ket{\bm{r}_2}e^{i(\bm{r}_2-\bm{r}_1)\cdot\bm{p}/\hbar}
\end{equation}

\noindent
where we have used $\braket{\bm{r}|\bm{p}}=e^{i\bm{r}\cdot \bm{p}/\hbar}/(2\pi\hbar)$ in two dimensions.  Terms in the integral for which $\bm{r}_1=\bm{r}_2$ are directly proportional to $n_r(\bm{r},t)$.

With the field $\bm{\mathcal{A}}^K$ defined above we can now calculate a {\it conjugate} topological invariant to $C_K$ we call the position space Chern number, $C_R$, as

\begin{widetext}
\begin{equation}
    C_R=\dfrac{V_{\text{cell}}}{2\pi}\sum_{nm} \dfrac{f_n-f_m}{(E_n-E_m)^2}\dfrac{i}{2}\bigg(\bra{n}\bm{\nabla}_{\mathcal{A}^k}\widehat{H}\ket{m}\times \bra{m}\bm{\nabla}_{\mathcal{A}^k}\widehat{H}\ket{n}\bigg) \cdot \bm{\hat{z}}
    \label{CRW}
\end{equation}
\end{widetext}

\noindent
with $V_{\text{cell}}=V/N$ is the volume per lattice site.  Like $C_K$, $C_R$ is integer valued in the thermodynamic limit at $T=0$ and is insensitive to perturbations that keep $E_n$ from crossing $\mu$.  In section \ref{CRCalc} we consider systems on a lattice with periodic boundary conditions and explain the underlying structure for which leads $C_R$ to take integer values.

\section{Calculating $C_R$ on a Lattice}\label{CRCalc}

Here we explore efficient ways to calculate $C_R$ for systems described in a non-interacting tight-binding framework.  For systems on periodic boundary conditions two natural bases to express the Hamiltonian, $\widehat{H}$, of the system are the positions basis $\ket{\bm{r}_i,\alpha}$ and momentum basis $\ket{\bm{k}_i,\alpha}$, where $\alpha=1,..,N_s$ with $N_s$ the number of degrees of freedom (or orbitals/spins) on a lattice site.  Here we consider a lattice of $N$ sites.  The basis states $\ket{\bm{r}_i,\alpha}$ are localized Wannier-like vectors satisfying $\widehat{\bm{r}}\ket{\bm{r}_i,\alpha}=\bm{r}_i\ket{\bm{r}_i,\alpha}$, with $\bm{r}_i=n_1\bm{R}_1+n_2\bm{R}_2$, for integers $n_d=1,..N_d$, and primitive  lattice vectors $\bm{R}_d$.  Their relationship to the momentum basis is given by 

\begin{equation}  \widehat{c}^\dagger_\alpha(\bm{r}_i)=\sum_{j=1}^N \dfrac{e^{-i\bm{r}_i\cdot\bm{k}_j}}{\sqrt{N}}\widehat{c}^\dagger_\alpha(\bm{k}_j)
\end{equation}

\noindent
with $\widehat{c}^\dagger_\alpha(\bm{r}_i)\ket{0}=\ket{\bm{r}_i,\alpha}$ and $\widehat{c}^\dagger_\alpha(\bm{k}_i)\ket{0}=\ket{\bm{k}_i,\alpha}$.  Periodic boundary conditions, $\bm{r}_i \sim \bm{r}_i+N_d\bm{R}_d$, restricts $\bm{k}_i$ to take values equal to $n_1\bm{b}_1/N_1+n_2\bm{b}_2/N_2$ with $\bm{b}_p$ reciprocal lattice vectors satisfying $\bm{b}_d\cdot\bm{R}_{d'}=2\pi\delta_{dd'}$.

The Hamiltonian of the system may be expressed in either basis

\begin{align}
\widehat{H}=\sum_{i,j=1}^N\sum_{\alpha,\beta=1}^{N_s}\widehat{c}^\dagger_\alpha(\bm{r}_i) t^R_{\alpha\beta}(\bm{r}_i,\bm{r}_j)c^{\phantom{\dagger}}_{\beta}(\bm{r}_j) \nonumber \\=\sum_{i,j=1}^N\sum_{\alpha,\beta=1}^{N_s}\widehat{c}^\dagger_\alpha(\bm{k}_i) t^K_{\alpha\beta}(\bm{k}_i,\bm{k}_j)c^{\phantom{\dagger}}_{\beta}(\bm{k}_j) 
\end{align}

\noindent
The relationship between the coupling functions $t^R_{\alpha\beta}(\bm{r}_i,\bm{r}_j)$ and $t^K_{\alpha\beta}(\bm{k}_i,\bm{k}_j)$ is given by a unitary transformation $U$ via $t^K_{\alpha\beta}=Ut^R_{\alpha\beta}U^\dagger$, with $U$, $t_{\alpha\beta}^R$, and $t_{\alpha\beta}^K$ matrices with $i,j$ coefficients $U_{ij}=e^{-i\bm{k}_i\cdot\bm{r}_j}/\sqrt{N}$, $t^R_{\alpha\beta}(\bm{r}_i,\bm{r}_j)$, and $t^K_{\alpha\beta}(\bm{k}_i,\bm{k}_j)$. 

In order to calculate $C_R$ we must first couple the system to $\bm{\mathcal{A}}^K$.  The field $\bm{\mathcal{A}}^K$ acts conjugately to $\bm{\mathcal{A}}^R$, which couples to the system via the Peierls substitution \cite{peierls1997theory,hofstadter1976energy}

\begin{equation} \widehat{c}^\dagger_\alpha(\bm{r}_i)c^{\phantom{\dagger}}_{\beta}(\bm{r}_j)\rightarrow \widehat{c}^\dagger_\alpha(\bm{r}_i)c^{\phantom{\dagger}}_{\beta}(\bm{r}_j)e^{-i(\bm{r}_i-\bm{r}_j)\cdot \bm{\mathcal{A}}^R}
\label{psubr}
\end{equation}

\noindent
Similarly, the field $\bm{\mathcal{A}}^K$ couples to the system via the substitution 

\begin{equation} \widehat{c}^\dagger_\alpha(\bm{k}_i)c^{\phantom{\dagger}}_{\beta}(\bm{k}_j)\rightarrow \widehat{c}^\dagger_\alpha(\bm{k}_i)c^{\phantom{\dagger}}_{\beta}(\bm{k}_j)e^{i(\bm{k}_i-\bm{k}_j)\cdot \bm{\mathcal{A}}^K}
\label{psubk}
\end{equation}

\noindent
Thse prescriptions are consistent with the substitutions $\widehat{\bm{p}}\rightarrow \widehat{\bm{p}}+\hbar \bm{\mathcal{A}}^R $ and $\widehat{\bm{r}}\rightarrow \widehat{\bm{r}}+\bm{\mathcal{A}}^K$ as can be most readily seen for systems with either lattice translation symmetry in position or momentum space.

If a system has lattice translation symmetry $\bm{r}_i\rightarrow \bm{r}_i+n_1\bm{R}_1+n_2\bm{R}_2$ then the coefficients $t^R_{\alpha\beta}(\bm{r}_i,\bm{r}_j)=t^R_{\alpha\beta}(\bm{r}_i-\bm{r}_j)$, and the Hamiltonian can be written in a Bloch form

\begin{equation}
\widehat{H}=\sum_{i=1}^N\sum_{\alpha,\beta=1}^{N_s}\widehat{c}^\dagger_\alpha(\bm{k}_i) H^K_{\alpha\beta}(\bm{k}_i)c^{\phantom{\dagger}}_{\beta}(\bm{k}_i) 
\end{equation}

\noindent
such that eigenstates of $\widehat{H}$, $\ket{u_n(\bm{k}_i)}$ are index by crystal momenta $\bm{k}_i$ and can be written as

\begin{equation}
    \ket{u_n(\bm{k}_i)}=\sum_\alpha \gamma_n^\alpha(\bm{k}_i)\widehat{c}^\dagger_\alpha(\bm{k}_i)\ket{0}
\end{equation}

\noindent
where the coefficients $\gamma_n^\alpha(\bm{k}_i)$ are determined by the eigenvectors of the matrix 

\begin{equation}
H^K_{\alpha\beta}(\bm{k}_i)=\sum_{\Delta_r}t_{\alpha\beta}^R(\bm{\Delta}_r)e^{-i\bm{\Delta}_r\cdot \bm{k}_i }
\end{equation}

\noindent
Therefore coupling the system to $\bm{\mathcal{A}}^R$ via \eqref{psubr} is equivalent to the substitution $H^K_{\alpha\beta}(\bm{k}_i)\rightarrow H^K_{\alpha\beta}(\bm{k}_i+\bm{\mathcal{A}}^R)$.

Likewise, for systems with momentum translation symmetry, $\bm{k}_i\rightarrow\bm{k}_i+n_1\bm{b}_1/N_1+n_2\bm{b}_2/N_2$, the coefficients $t^K_{\alpha\beta}(\bm{k}_i,\bm{k}_j)=t^K_{\alpha\beta}(\bm{k}_i-\bm{k}_j)$, and the Hamiltonian can be written in a local form

\begin{equation}
\widehat{H}=\sum_{i=1}^N\sum_{\alpha,\beta=1}^{N_s}\widehat{c}^\dagger_\alpha(\bm{r}_i) H^R_{\alpha\beta}(\bm{r}_i)c^{\phantom{\dagger}}_{\beta}(\bm{r}_i) 
\end{equation}

\noindent
such that eigenstates of $\widehat{H}$, $\ket{w_n(\bm{r}_i)}$ are index by a lattice site $\bm{r}_i$ and can be written as

\begin{equation}
    \ket{w_n(\bm{r}_i)}=\sum_\alpha \xi_n^\alpha(\bm{r}_i)\widehat{c}^\dagger_\alpha(\bm{r}_i)\ket{0}
\end{equation}

\noindent
where the coefficients $\xi_n^\alpha(\bm{r}_i)$ are determined by the eigenvectors of the matrix 

\begin{equation}
H^R_{\alpha\beta}(\bm{r}_i)=\sum_{\Delta_k}t_{\alpha\beta}^R(\bm{\Delta}_k)e^{i\bm{\Delta}_k\cdot \bm{r}_i }
\end{equation}

\noindent
Therefore coupling the system to $\bm{\mathcal{A}}^K$ via \eqref{psubk} is equivalent to the substitution $H^R_{\alpha\beta}(\bm{r}_i)\rightarrow H^R_{\alpha\beta}(\bm{r}_i+\bm{\mathcal{A}}^K)$.

The quantization of $C_R$ and $C_K$ can be understood as follows.  In the context of the momentum space Chern insulator the correlation function in \eqref{CKG} describes the response of the system's eigenstates as one threads an Aharonov-Bohm-like flux through the system, which amounts to putting a twisted boundary condition on the state ket $\widehat{T}^R_{N_d\bm{R}_d}\ket{\Psi}=e^{i\phi^R_d}\ket{\Psi}$, where $\widehat{T}^R_{N_d\bm{R}_d}$ translates the positions of the system along one of its two distinct periodic directions \cite{laughlin1981quantized,thouless1982quantized}.  Similarly equation \eqref{CRW} describes how a system responds to the application of a flux, that twists the boundary condition of the wavefunction on the momentum space torus $\widehat{T}_{\bm{b}_d}\ket{\Psi}=e^{i\phi^K_d}\ket{\Psi}$, where $\widehat{T}^K_{\bm{b}_d}$ translates the system's momenta along one of its two distinct periodic directions by $\bm{b}_d$.  The boundary conditions provide means of constructing two different Berry connections $\bm{\omega}^R=i\bra{\Psi}\bm{\nabla}_{\phi^R}\ket{\Psi}$ and $\bm{\omega}^K=i\bra{\Psi}\bm{\nabla}_{\phi^K}\ket{\Psi}$, where $\bm{\phi}^R=\phi^R_1\bm{\hat{x}}+\phi^R_2\bm{\hat{y}}$ and $\bm{\phi}^K=\phi^K_1\bm{\hat{x}}+\phi^K_2\bm{\hat{y}}$, and curvatures $\bm{F}^R=\bm{\nabla}_{\phi^R}\times \bm{\omega}^R$ and $\bm{F}^K=\bm{\nabla}_{\phi^K}\times \bm{\omega}^K$.  The momentum space and position space Chern numbers are given by

\begin{align}
    C_K=\dfrac{1}{2\pi}\int d^2\phi^R \bm{F}^R\cdot{\bm{\hat{z}}} \,\,,
    C_R=\dfrac{1}{2\pi}\int d^2\phi^K \bm{F}^K\cdot{\bm{\hat{z}}}
\end{align}

\noindent
which reproduce equations \eqref{CKG} and \eqref{CRW} when the equations are evaluated in the thermodynamic limit \cite{laughlin1981quantized,niu1985quantized}.

\section{Skyrmion Magnetic Phases}
\label{CaN}

Here we explore some simple systems that exhibit nontrivial position space topology with $C_R\neq 0$.  We take a model of conduction electrons coupled to a local skyrmion magnetic texture, $\bm{m}(\bm{r})$.  Consider a 2D square lattice of size $L\times L$ and lattice constant $a$ with Hamiltonian $\widehat{H}=\widehat{H}_T+\widehat{H}_R$.  We couple nearest neighbor sites on the lattice such that the kinetic contribution to $\widehat{H}$, $\widehat{H}_T$, takes the form

\begin{equation}
    \widehat{H}_T=t \sum_{i=1}^N\sum_{\alpha=\uparrow,\downarrow} 2(\cos(k_x a)+\cos(k_y a))\widehat{c}^\dagger_{\alpha}(\bm{k}_i)\widehat{c}^{\phantom{\dagger}}_{\alpha}(\bm{k}_i)
\end{equation}

\noindent
Here $\alpha$ corresponds to the internal spin degree of freedom of the electrons on each site.  The magnetic texture locally couples to the spin degrees of freedom of electrons on the lattice via $\widehat{H}_R$, which takes the form

\begin{equation}
\widehat{H}_R=t_R\sum_{i=1}^N\sum_{\alpha,\beta=\uparrow,\downarrow}\widehat{c}^\dagger_{\alpha}(\bm{r}_i)(\bm{m}(\bm{r}_i)\cdot\bm{\sigma}_{\alpha\beta})\widehat{c}^{\phantom{\dagger}}_\beta(\bm{r}_i)
\label{HSK}
\end{equation}

\noindent
with $\bm{\sigma}=\sigma_x\bm{\hat{x}}+\sigma_x\bm{\hat{y}}+\sigma_x\bm{\hat{z}}$, and $\sigma_i$ the Pauli matrices.  The constants $t$ and $t_R$ set the ratio of the coupling strength between nearest neighbor interactions and the magnetic texture.  Here we choose $\bm{m}(\bm{r})=\sin(2\pi x_i/L)\bm{\hat{x}}+\sin(2\pi y_i/L)\bm{\hat{y}}+(\cos(2\pi x_i/L)+\cos(2\pi y_i/L)-\eta_R)\bm{\hat{z}}$, with $\bm{r}_i=x_i\bm{\hat{x}}+y_i\bm{\hat{y}}$.

\begin{figure*}
 \begin{centering}
\includegraphics[width=.95\textwidth]{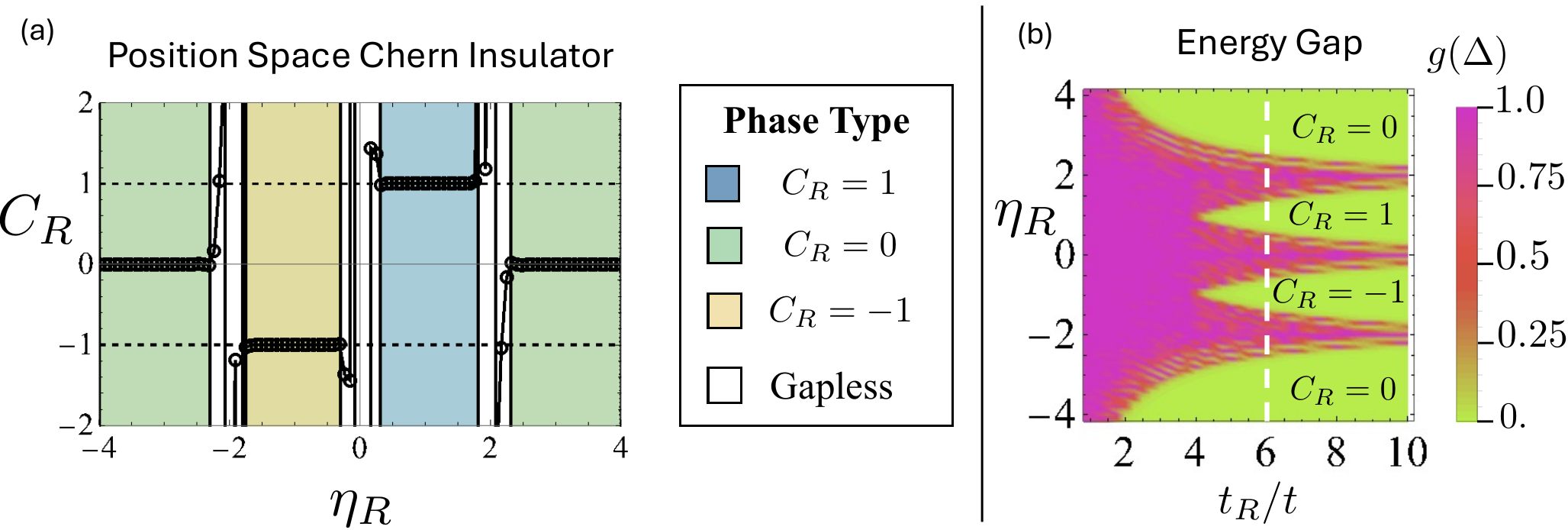}
\caption{{\bf Position Space Chern Number:}  (a) Evolution of $C_R$ as a function of $\eta_R$ for a system $\widehat{H}=\widehat{H}_T+\widehat{H}_R$ with $t_R=6t$ and $L=30a$ at half filling and $T=0$.  Topological phase transitions occur around $\eta_R=0,\pm 2$ where the system becomes gapless.  (b) Energy gap as a function of $\eta_R$ and the ratio of  the coupling $t_R/t$.  Topologically distinct phases are separated by gapless regions.  Dashed white line represents the phase space cut shown in (a).}
 \label{CrPDis}
 \end{centering}
\end{figure*}

Figure \ref{CrPDis}(a) shows the position space Chern number, $C_R$, as a function of $\eta_R$ for a system with $t_R=6t$, with $L=30a$, at half filling, and $T=0$.  Topological transitions occur around $\eta_R=0,\pm2$ where the system becomes gapless.  The value of $C_R$ can be understood as follows.  In the limit $t\rightarrow 0$ the system becomes translationally invariant in momentum space such that eigenstates of the Hamiltonian take the form of $\ket{w_n(\bm{r}_i)}$ with $\xi_+^\uparrow(\bm{r})= \cos(\theta_i/2),\,\,\xi_+^\downarrow(\bm{r}_i)=e^{i\phi_i}\sin(\theta_i/2),\,\, \xi_-^\uparrow(\bm{r}_i)= e^{-i\phi_i}\sin(\theta_i/2),\,$ and $\,\xi_-^\downarrow(\bm{r})=-\cos(\theta_i/2)$.  Here $(\phi_i,\theta_i)$ are the spherical and polar angles parameterizing $\hat{\bm{m}}(\bm{r}_i)$.  Substitution into \eqref{CRW} and taking the thermodynamic limit leads to $C_R=-N_{\text{sk}}$ (see equation \eqref{NSK}) \cite{bernevig2013topological}.  For the chosen $\bm{m}(\bm{r})$ above

\begin{equation}
    N_{\text{sk}}=\begin{cases}
        -\text{sgn}(\eta_R) & -2<\eta_R<2 \\
        0 & |\eta_R|>2
    \end{cases}
    \label{etaSK}
\end{equation}

Even for system with $t\neq 0$ that break translation symmetries in momentum space, the Chern number reflects the classical winding number of the vector field $\hat{\bm{m}}(\bm{r})$ as the topological index, $C_R$, is insensitive to perturbations of the system that keep states from crossing the chemical potential.  For phases $C_R\neq 0$ and small $t$ there is a smooth deformation of the spectrum to the $t=0$ limit without closing the energy gap and the topology of the system remains invariant.  Figure \ref{CrPDis}(b) shows the energy gap $\Delta$ at half filling as a function of $\eta_R$ and $t_R/t$.  As $t$ approaches $t_R/2$ the system becomes gapless for a large part of the configuration space and the winding of $\hat{\bm{m}}(\bm{r})$ no longer determines the position space topology of the system.  In order to better visualize gap closures, the function $g(\Delta)=(1+(4\Delta/t)^2)^{-1}$ is plotted such that gapless phases occur when $g(\Delta)\sim 1$.  Distinct topological phases are separated by gapless regions.  A white dashed line marks the phase space cut shown in Fig \ref{CrPDis}(a).

\section{Bulk-Edge Correspondence for $C_R$}

One of the hallmarks of Chern insulators is that they exhibit a bulk-edge correspondence \cite{halperin1982quantized,wen1990chiral}.  In the context of the momentum space Chern insulator, periodic systems for which $C_K\neq 0$ exhibit states within the bulk insulating energy gap that are localized to a position space boundary when put into a ribbon geometry.  Here we explore the consequences of $C_R\neq 0$ by investigating the effects of this spatial constraint on the skyrmion magnetic phases described above.

We consider a position space Chern insulator in a ribbon geometry where we impose periodic boundary conditions in the $x$-direction of the ribbon, but couple the system to an edge potential in the $y$-direction.  In order to see the unique bulk-edge correspondence and associated {\it edge}-like states for these systems, we must also couple the system to an {\it edge} that acts like a potential barrier in momentum space.  First we demonstrate this phenomena using the canonical description of an {\it edge}.  We then proceed to try and engineer an {\it edge} potential that is most suitable for application in a real experimental setting.

To determine how to implement a momentum space {\it edge} we must first express $\widehat{H}_R$ in the momentum basis.  The magnetic texture $\bm{m}(\bm{r})$ couples to conduction electrons in $\widehat{H}_R$ in position space through an onsite interaction.  The couplings oscillate with a wavevector $\sim L/2\pi$.  On the momentum space lattice this corresponds to a nearest neighbor hopping between sites whose momenta differ by $2\pi/L$.  To implement a momentum space {\it edge} we choose to set to zero the coupling between states at the extrema of the momentum space lattice.  To visualize and understand physically the effect of this type of {\it edge} on $\widehat{H}_R$, we can then transform the modified couplings back to the position basis.

To demonstrate how this type of {\it edge} effects the couplings in $\widehat{H}$, for simplicity consider a finite 1D chain of length $L$ with $N$ odd sites located at positions $x_n=na$, $n=1,..,N$, with nearest neighboring coupling of strength $t_c$ on the momentum space lattice:

\begin{equation}
    \widehat{H}_{1D}=t_c\sum^{N-1}_{i=1}\bigg(\widehat{c}^\dagger(k_i)\widehat{c}(k_{i+1})+\widehat{c}^\dagger(k_{i+1})\widehat{c}(k_i)\bigg)
\end{equation}

\begin{figure}
\includegraphics[width=.48\textwidth]{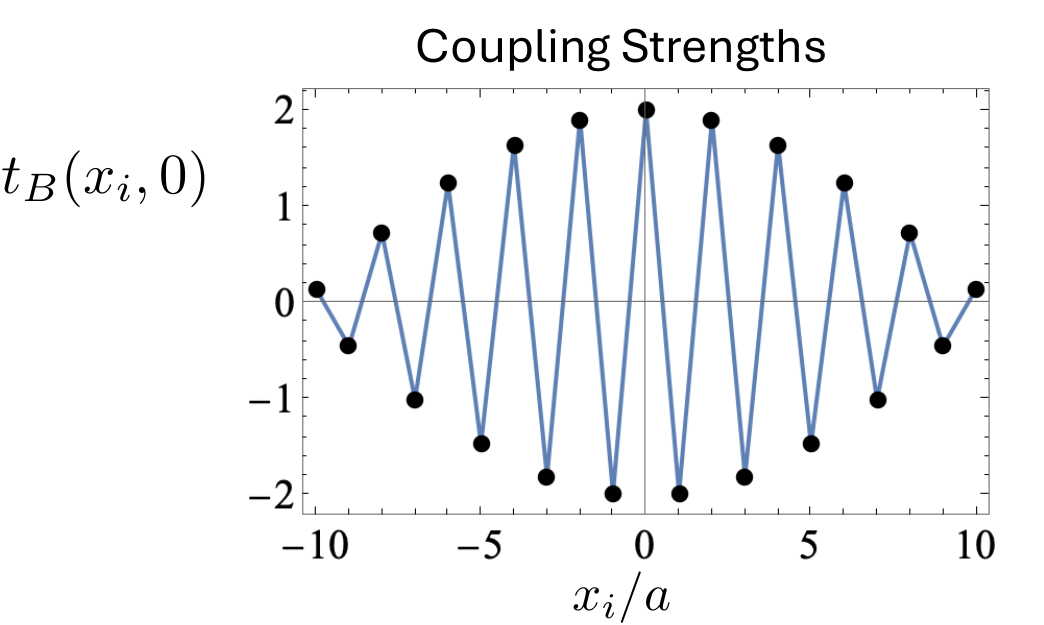}
\caption{{\bf 1D Momentum Space \textit{\textbf{Edge}} Potential:}  Coupling strengths $t_B(x_i,0)$ for a site in the middle of the chain of size $N=21$.  Effect of boundary is to induce coupling between sites that oscillate at wavelengths $\sim 2a, 2L$.}
 \label{boundcoup}
\end{figure}

\noindent
Here we choose $k_n=(2n-N-1)\pi/L$ such that the center of the momentum space lattice is at the origin.  If the system was on periodic boundary conditions the sum would go from $i$ equals $1$ to $N$ (rather than $N-1$), with the $N+1$ site being identified with the site on the chain at $N=1$, and the system would be invariant under translations $k_i\rightarrow k_i+2\pi n/L$, for integers $n$.  In this translationally periodic limit the energy eigenstates would be localized states, $\ket{w(x_i)}$, indexed by a lattice site $x_i$ and having eigenenergies $E(x_i)=2t_c\cos(2\pi x_i/L)$.  For the finite chain the momentum space translation symmetry is broken and $\widehat{H}_{1D}$ couples orbitals on the position space lattice localized at different sites.  This can be seen by writing $\widehat{H}_{1D}=\widehat{H}_{PBC}-\widehat{H}_{B}$, with

\begin{equation}
\widehat{H}_{PBC}=\sum_{i=1}^N\widehat{c}^\dagger(x_i)E(x_i)\widehat{c}(x_i)
\end{equation}

\noindent
and 

\begin{align}
\widehat{H}_{B}&=t_c\bigg(\widehat{c}^\dagger(k_N)\widehat{c}(k_{1})+\widehat{c}^\dagger(k_{1})\widehat{c}(k_N)\bigg) \nonumber\\
&=\dfrac{1}{N}\sum_{i,j=1}^N\widehat{c}^\dagger(x_i)t_B(x_i,x_j)\widehat{c}(x_j)
\end{align}

\noindent
where $t_B(x_i,x_j)=2t_c\cos((N-1)\pi(x_i+x_j)/L)$. $\widehat{H}_{PBC}$ describes the fully periodic chain and $\widehat{H}_{B}$ describes the effects of the momentum space {\it edge} potential on the ribbon.  The effect of the {\it edge} is to introduce long range coupling between sites that oscillates at wavelengths $\sim 2a$ and $2L$.  Figure \ref{boundcoup} shows the position space couplings, $t_B(x_i,0)$, induced by this type of momentum space wall.

We now consider a position space Chern insulator described by $\widehat{H}=\widehat{H}_T+\widehat{H}_R$, but in a ribbon geometry in position space, with a total odd number of sites, $N$.  To observe the {\it edge} states associated to the position space topology of the bulk, we couple the system to a momentum dependent potential by setting to zero the coupling between states with momenta, $k_y$, at the extrema of allowed values.  This is analogous to the 1D finite chain described above where the couplings between the states $\widehat{c}^\dagger(k_1)\ket{0}$ and  $\widehat{c}^\dagger(k_N)\ket{0}$ are set to zero.  The Hamiltonian corresponding to the system in this ribbon geometry, $\widehat{H}_{\text{RG}}$ can be written as

\begin{equation}
\widehat{H}_{\text{RG}}=\widehat{H}_T+\widehat{H}_R-\widehat{H}_{B_T}-\widehat{H}_{B_R}
\end{equation}

\noindent
where

\begin{widetext}
\begin{align}
    \widehat{H}_{B_T}&=\sum_{\bm{k}_i \in \bm{K}^*_1}\,\sum_{\bm{k}_j\in \bm{K}^*_2}\sum_{\alpha,\beta=\uparrow,\downarrow}\bigg(\bra{\bm{k}_i,\alpha}\widehat{H}_T\ket{\bm{k}_j,\beta}+\bra{\bm{k}_i,\alpha}\widehat{H}_R\ket{\bm{k}_j,\beta}\bigg)\widehat{c}^\dagger_\alpha(\bm{k}_i)\widehat{c}^{\phantom{\dagger}}_\beta(\bm{k}_j) +(i\leftrightarrow j) \\
     \widehat{H}_{B_R}&=\sum_{\bm{r}_i \in \bm{R}^*_1}\,\sum_{\bm{r}_j\in \bm{R}^*_2}\sum_{\alpha,\beta=\uparrow,\downarrow}\bigg(\bra{\bm{r}_i,\alpha}\widehat{H}_T\ket{\bm{r}_j,\beta}+\bra{\bm{r}_i,\alpha}\widehat{H}_R\ket{\bm{r}_j,\beta}\bigg)\widehat{c}^\dagger_\alpha(\bm{r}_i)\widehat{c}^{\phantom{\dagger}}_\beta(\bm{r}_j)+(i\leftrightarrow j)
\end{align}
\end{widetext}

\noindent
with $\bm{K}_1^*$ and $\bm{K}_2^*$ the collection of $\bm{k}_i$ with $\bm{k}_i\cdot \bm{\hat{y}}=(1-N)\pi/L$ and $\bm{k}_i\cdot \bm{\hat{y}}=(N-1)\pi/L$ respectively, and $\bm{R}_1^*$ and $\bm{R}_2^*$ the collection of $\bm{r}_i$ with $\bm{r}_i\cdot \bm{\hat{y}}=a$ and $\bm{r}_i\cdot \bm{\hat{y}}=L$ respectively.  For this model system, we note that $\bra{\bm{k}_i,\alpha}\widehat{H}_T\ket{\bm{k}_j,\beta}\sim\delta_{ij}$ and $\bra{\bm{r}_i,\alpha}\widehat{H}_R\ket{\bm{r}_j,\beta}\sim\delta_{ij}$, such that these terms do not contribute to $\widehat{H}_{B_T}$ or $\widehat{H}_{B_R}$.  The edge potential $\widehat{H}_{B_R}$ represents the couplings induced by the physical position space boundary associated with the real space edge of the sample, while $\widehat{H}_{B_T}$ can be thought of as representing some momentum dependent perturbation to the system. 

\begin{figure*}
 \begin{centering}
\includegraphics[width=.95\textwidth]{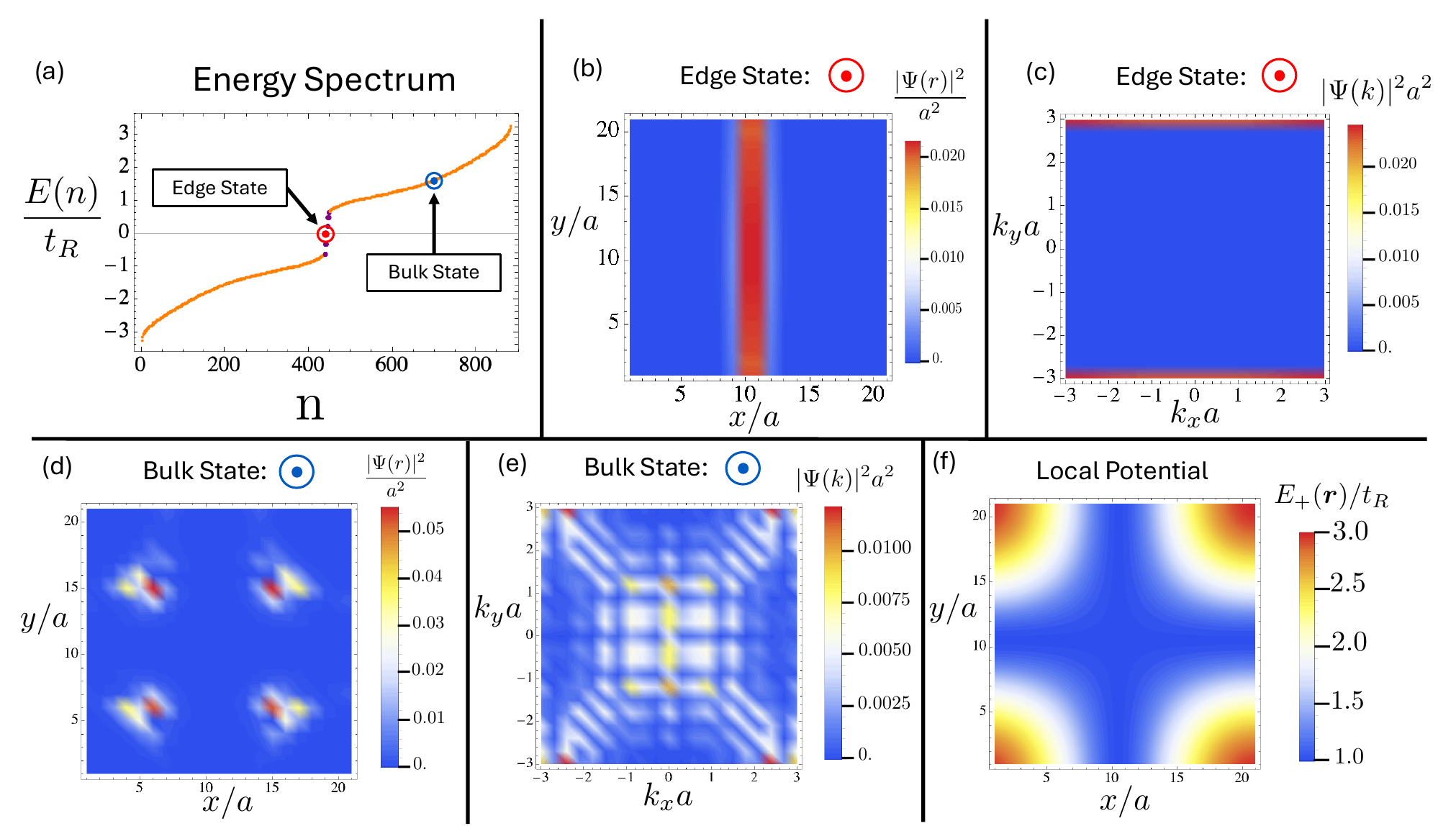}
\caption{{\bf Bulk-edge Correspondence for Position Space Chern Insulators:}  Here we consider a system, $\widehat{H}_{\text{RG}}$, with $N=21^2=441$ sites and $\eta_R=-1$.  (a) Energy spectrum with $t=0.1t_R$ consisting of bulk states (orange) and {\it edge} states (purple). (b)-(c)  Local probability density in the position (b), $|\Psi(\bm{r})|^2$, and momentum (c), $|\Psi(\bm{k})|^2$, basis for {\it edge} state highlighted (red) in (a).  (d)-(e) Local probability density in the position (d), $|\Psi(\bm{r})|^2$, and momentum (e), $|\Psi(\bm{k})|^2$, basis for typical bulk state highlighted (blue) in (a). (f) Local potential $E_+(\bm{r})=t_R|\bm{m}(\bm{r})|$ showing minima at center of ribbon where low energy {\it edge} states traverse the ribbon.}
 \label{BECor}
 \end{centering}
\end{figure*}

Figure \ref{BECor} shows a system with $N=21^2=441$ sites with $t=0.1t_R$ and $\eta_R=-1$.  The energy spectrum is shown in figure \ref{BECor}(a) showing bulk (orange) and {\it edge} states (purple) that traverse the energy gap.  Figure \ref{BECor}(b)-(c) show the local probability density for the {\it edge} state highlighted (red) in  figure \ref{BECor}(a) in the position basis (b), $|\Psi(\bm{r})|^2$, and momentum basis (c), $|\Psi(\bm{k})|^2$.  In the momentum basis, $|\Psi(\bm{k})|^2$ is localized at the extrema of the momentum space lattice.  In position space this corresponds to a state that traverses the extent of the ribbon along its finite direction and is centered at a position where states near the chemical potential in the absence of the position and momentum space walls are localized.  To see this we can look at the $t=0$ limit for which the coupling in $\widehat{H}$ becomes purely local in position space.  Eigenenergies then take the form of $E_\pm(\bm{r}_i)=\pm t_R |\bm{m}(\bm{r}_i)|$, (Fig. \ref{BECor}(f)).  The {\it edge} states have energies within the bulk energy gap and are thus centered near where $E_+$ ($E_-$) takes its minimum (maximum) values.  For this particular magnetic texture, $\eta_R=-1$, minima of $E_+$ occur near the center of the ribbon and the {\it edge} states closest to $E=0$ are localized here (see Fig. \ref{BECor}(b)).

\subsection{Other Perturbations For Observing {\it Edge} States}

The conventional {\it edge} structure in momentum space leads to long range couplings between sites on the position space lattice that may be challenging to implement in an experimental setting.  The {\it edge states} associated with the bulk-boundary correspondence endowed by the position space Chern number will exist as long as there is a momentum space potential in the system that acts like an {\it edge}, i.e. has a magnitude that traverses the bulk energy gap and with an extent that keeps the edge states from hybridizing.  Here we investigate a class of gaussian potentials as alternative perturbations that demonstrate a bulk-boundary correspondence and do not rely on the long range coupling structure induced by the canonical {\it edge} potential.

Consider a 1D system on periodic boundary conditions subjected to the potential

\begin{equation}
    \widehat{H}_{w}^{1D}=\sum_i \widehat{c}^\dagger(k_i)V^{1D}_k(k_i)\widehat{c}(k_{i})
\end{equation}

\begin{figure}
 \begin{centering}
\includegraphics[width=.48\textwidth]{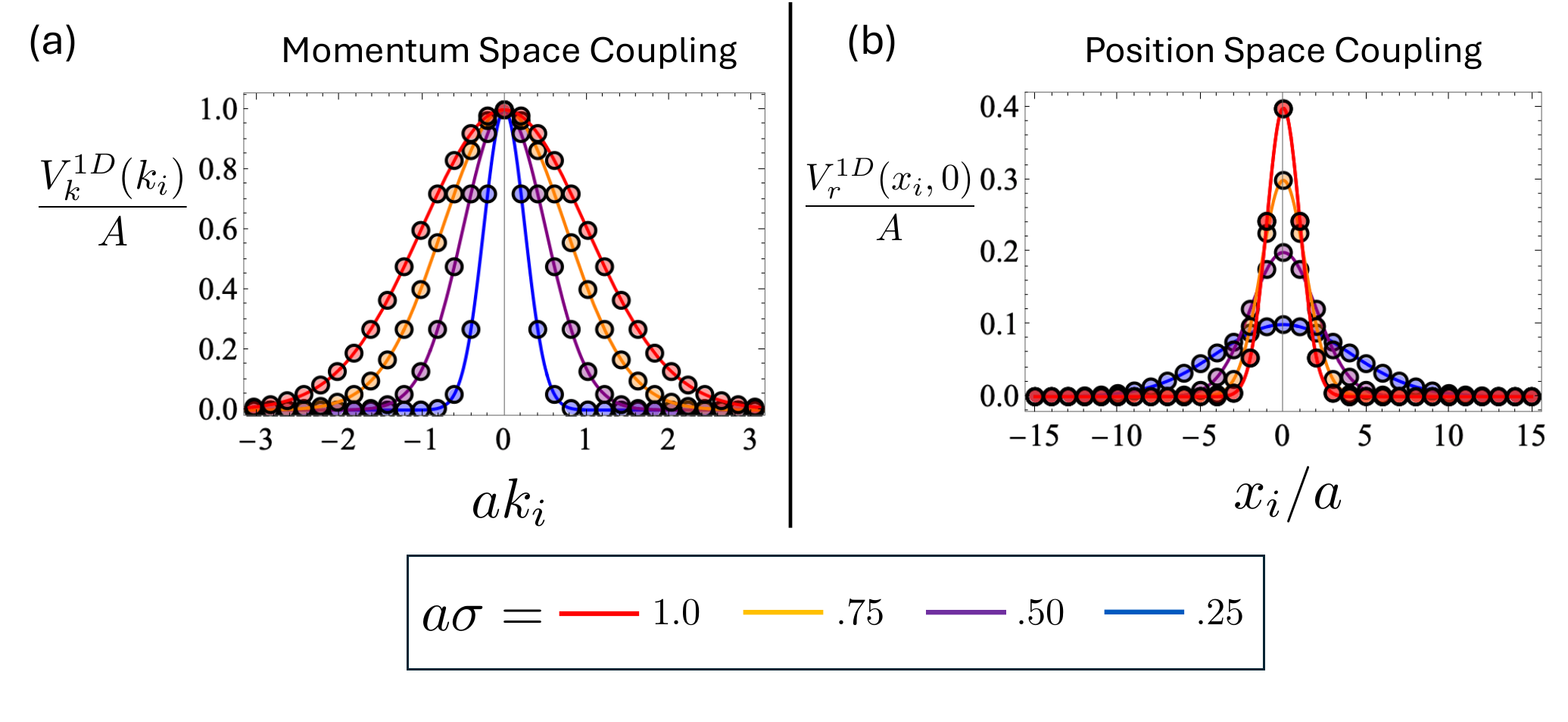}
\caption{{\bf Coupling Strengths for 1D Gaussian Potential:}  In order to make the position space couplings induced by a momentum space wall $\widehat{H}^{1D}_w$ local, we may consider a wide gaussian distribution $V^{1D}_k(k)$ in momentum space. (a) Coupling coefficients in momentum space for gaussian wall of various widths for a 1D chain of $N=31$ sites with lattice constant $a$. (b)  Corresponding position space couplings $V^{1D}_r(x_i,0)$.  As the momentum space wall width increases $V^{1D}_r(x_i,0)$ becomes more local.}
 \label{gausswall}
 \end{centering}
\end{figure}

\noindent
Here we take $V^{1D}_k(k_i)=Ae^{-k_i^2/2\sigma^2}$ to take a gaussian form parameterized by its width $\sigma$ and its amplitude at $k_i=0$, $A$.  In order to maintain well defined periodic boundary conditions we take $\sigma$ such that $V^{1D}(k_i)\rightarrow 0$ near the extrema of the momentum space lattice.  In position space this potential can be written as

\begin{equation}
    \widehat{H}_{w}^{1D}=\sum_{ij} \widehat{c}^\dagger(x_i)V^{1D}_r(x_i,x_j)\widehat{c}(x_{j})
\end{equation}

\noindent
with $V^{1D}_r(x_i,x_j)\approx \dfrac{A\sigma}{\sqrt{2\pi}}e^{-\sigma^2(x_i-x_j)^2/2}$.  Figure \ref{gausswall} shows the coupling strength for a 1D gaussian potential barrier in momentum (a) and position (b) space. For this potential the interaction between position space lattice sites is short range, with the characteristic length scale describing the interactions decreasing with increasing $\sigma$.

\begin{figure}
 \begin{centering}
\includegraphics[width=.48\textwidth]{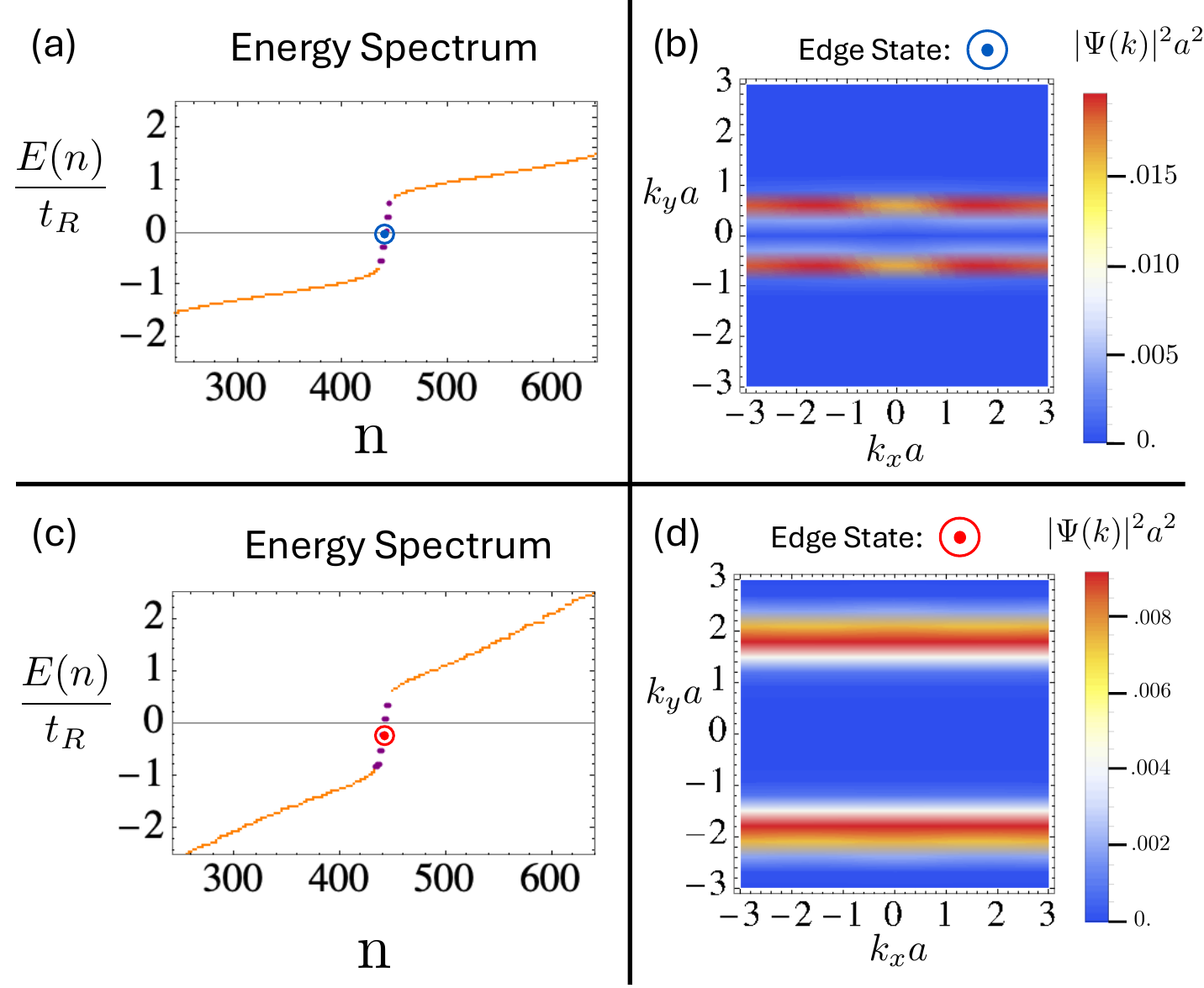}
\caption{{\bf Edge states for Gaussian Potential Wall:}  Energy spectrum and representative edge state for a system with $\widehat{H}=\widehat{H}_T+\widehat{H}_R$ in a ribbon geometry with periodic boundary conditions in the $x$-direction, but of finite extent in the $y$-direction with $t=0.1t_R$ and $\eta_R=-1$.  Ribbon contains $N=21^2$ sites with lattice constant $a$ subjected to a gaussian potential with $A=4.5t_R$ and $a\sigma=0.25$ ((a)-(b)) and $a\sigma=1$ ((c)-(d)).  Edgestates are localized near $\partial^2_{k_y}V(k_y)=0$.}
 \label{ESgauswall}
 \end{centering}
\end{figure}

To demonstrate the bulk-boundary correspondence using this type of gaussian wall we again consider a system $\widehat{H}=\widehat{H}_T+\widehat{H}_R$ in the same position space ribbon geometry given above.  We then couple this system to a gaussian wall:

\begin{equation}
\widehat{H}_w=\sum_{i,\alpha=\uparrow\downarrow}\widehat{c}_\alpha^\dagger(\bm{k}_i)s(\alpha) V(k_{y_i})\widehat{c}_\alpha^{\phantom{\dagger}}(\bm{k}_i)
\end{equation}

\noindent
with $V(k_{y_i})=Ae^{-k^2_{y_i}/2\sigma^2}$ and $s(\alpha)$ equal $+1$ for spin up and $-1$ for spin down. For sufficiently large $A$, $\widehat{H}_w$ produces a spin dependent potential at the center of the wall that locally in momentum space can be associated with a trivial insulator.  This is tantamount to taking $\eta_R\rightarrow \eta_R(k_y)$ such that a portion of the wall takes values where $\eta_R<-2$ (see Fig. \ref{CrPDis}(a)), introducing into the sample a region where the system is topologically trivial.  Figure \ref{ESgauswall} shows the energy spectrum for a system with $N=21^2$ sites and lattice constant $a$, with $t=0.1t_R$ and $\eta_R=-1$.  The potential wall is chosen to have $A=4.5t_R$ and widths $\sigma=0.25/a$, ((a)-(b)), and  $\sigma=1/a$, ((c)-(d)).  Energy spectrum plots show bulk states (orange) and {\it edge} states (purple).  A representative {\it edge} state for each wall is plotted in (c) and in (d) showing localization of the {\it edge} modes along either side of potential barrier.  Edgesates are localized near $\partial_{k_y}^2V(k_y)=0$.

\subsection{Trivial Insulators}

\begin{figure}
 \begin{centering}
\includegraphics[width=.48\textwidth]{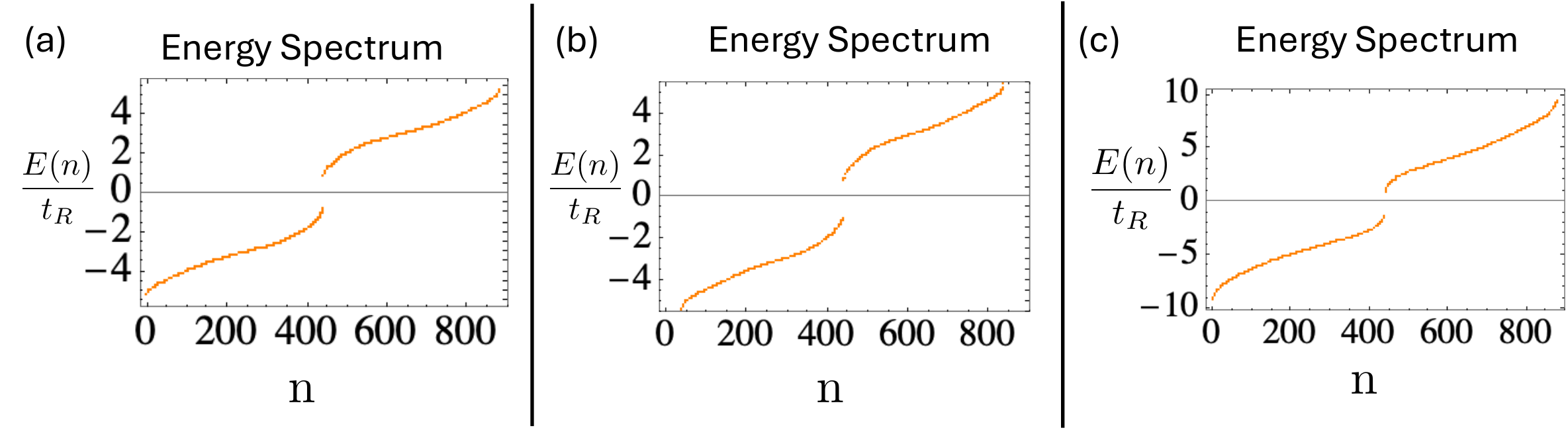}
\caption{{\bf Trivial Insulators:}  Energy spectra for systems shown in Fig. \ref{BECor}(a), \ref{ESgauswall}(a), and \ref{ESgauswall}(b) (shown in Fig. \ref{trivial} (a), (b), and (c) respectively), with the same system parameters, but $\eta_R=-3$.}
 \label{trivial}
 \end{centering}
\end{figure}

For systems $\widehat{H}=\widehat{H}_T+\widehat{H}_R$ the topology is dependent on the value of $\eta_R$.  As shown in figure \ref{CrPDis}(b) the phase space for which $C_R=0$, occur away from gapless regions where $|\eta_R|\gtrsim2$.  Fig. \ref{trivial} shows the corresponding trivial insulating energy spectrum for systems shown in Fig. \ref{BECor}(a), \ref{ESgauswall}(a), and \ref{ESgauswall}(b), with the same parameters, but $\eta_R=-3$.  In this regime the systems have a a bulk trivial topology, and as such no states cross the chemical potential at $\mu=0$ and only bulk states are present in the systems' spectra.

\section{Physical Observables Associated with $C_R$}

Above we have highlighted the bulk-boundary correspondence of position space Chern insulators, presenting a pathway to measuring the topology of these systems through the observation of its topological {\it edge} states when the system is perturbed by a momentum space wall-like potential.  Like for momentum space Chern insulators, position space topology also indicates properties that manifest in the bulk of the sample.  

In the case of momentum space Chern insulators, $C_K\neq 0$ leads to a bulk Hall conductivity $\sigma_H$ that describes the transportation of charge across the system, perpendicular to a locally applied electric field.  As discussed in section \ref{PSCN} the relations $j_i^Q=\sum_{j=x,y}\sigma_{ij}E_j$ and $C_K=h\sigma_H/e^2$ lead to equation \eqref{CKG}: an expression for $C_K$ in the form of a transverse current-current correlation function.  The effects of a nonzero Hall conductivity (or $C_K$) manifest in a Chern-Simons term in the effective action of the system of the form 

\begin{equation}
    S_{\text{eff}}\sim \dfrac{\sigma_H}{2}\sum_{\mu,\nu,\rho=0}^2\int d^2r dt \,\varepsilon^{\mu\nu\rho}A_\mu\partial_\nu A_\rho
\end{equation}

\noindent
with $A_\mu=(-\phi/c,A_x,A_y)$ whose equations of motion reproduce $j_i^Q=\sum_j \sigma_H \epsilon^{ij}E_j$ and the relation $\partial_t n_Q=\sigma_H\partial_tB$ due to charge conservation, where $n_Q$ is the local charge density (see equation \eqref{conR}) and $\phi$ the electromagnetic scalar potential \cite{arovas1985statistical,qi2008topological}.

In the case of the position space Chern insulator, equation \eqref{CRW} expresses $C_R$ in terms of correlations between the components of the operator $\bm{\nabla}_{\mathcal{A}^K}\widehat{H}$ that describe the flux of particles in momentum space (or the force density on charge carriers, see section \ref{PSCN}). As such the linear response of a system to a constant $\bm{V}=-\partial_t \bm{\mathcal{A}}^K(t)$ is described through the relation $j^{p}_i=\sum_{j=x,y}\chi_{ij} V_j$ with $\chi_H=(\chi_{xy}-\chi_{yx})/2$ and $\chi_H=C_R/2\pi\hbar$.  Similiarly to the Hall conductivity, nonzero $\chi_H$, manifests in a term in the effective action of the form

\begin{equation}
    S\sim \dfrac{\chi_H}{2}\sum_{\mu,\nu,\rho=0}^2\int d^2k dt \,\varepsilon^{\mu\nu\rho}\mathcal{A}^K_\mu\partial^{\phantom{s}}_\nu \mathcal{A}^K_\rho
\end{equation}

\noindent
with $\mathcal{A}^K_\mu=(-c\Gamma ,\mathcal{A}^K_x,\mathcal{A}^K_y)$ whose equations of motion, for $\Gamma=0$, reproduce $j_i^p=\sum_j \chi_H \epsilon^{ij}V_j$ and the relation $\partial_t n_p=\chi_H\partial_t(\bm{\nabla}_p \times \bm{\mathcal{A}}^K)\cdot \bm{\hat{z}}$.  The exploration of perturbations with non-zero time component, $\mathcal{A}^K_0=-c\Gamma$ is left for future work.

The field $\bm{\mathcal{A}}^K$ couples to the system via the substitution $\bm{\hat{r}}\rightarrow \bm{\hat{r}}+\bm{\mathcal{A}}^K$.  A time dependent field $\bm{\mathcal{A}}^K(t)$ describes a flow of electrons in the system.  This is most readily understood in the context of semiclassical dynamics. For systems with position space translation symmetry subjected to a long wavelength local perturbation a semiclassical theory can be defined.  Following \cite{sundaram1999wave,xiao2005berry,xiao2010berry,freimuth2013phase} a position space gradient expansion of $\widehat{H}$ leads to a semiclassical Hamiltonian 

\begin{equation}
    \widehat{H}^{\text{sc}}(\bm{r},\bm{k})= \sum_{\alpha,\beta=1}^{N_s} \widehat{c}^\dagger_\alpha(\bm{r},\bm{k}) t^{\text{sc}}_{\alpha\beta}(\bm{r},\bm{k})\widehat{c}^{\phantom{\dagger}}_\beta(\bm{r},\bm{k})
\end{equation} 

\noindent
where $t^{\text{sc}}_{\alpha\beta}(\bm{r},\bm{k})=\bra{W(\bm{r},\bm{k},\alpha)}\widehat{H}\ket{W(\bm{r},\bm{k},\beta)}$, with $c^\dagger_\alpha(\bm{r},\bm{k})\ket{0}=\ket{W(\bm{r},\bm{k},\alpha)}$ a localized wavepacket with orbital character $\alpha$, and an average crystal momentum $\hbar \bm{k}$ and position $\bm{r}$.  The eigenenergies and eigenstates of $\widehat{H}^{\text{sc}}(\bm{r},\bm{k})$ we denote as $\mathcal{E}_n(\bm{r},\bm{k})$ and $\ket{W_n(\bm{r},\bm{k})}$.

The semiclassical equations of motion describing the time evolution of the phase space variables $\bm{r}$ and $\bm{k}$ for wavepacket $\ket{W_n(\bm{r},\bm{k})}$ can be written as 

\begin{align}
    \pdv{\bm{r}}{t} &= \dfrac{1}{ \hbar\mathcal{D}}\bigg[  \overleftrightarrow{\mathbf{K}}\bm{\nabla}_r \widetilde{\mathcal{E}}+\overleftrightarrow{\mathbf{S}} \bm{\nabla}_k\widetilde{\mathcal{E}} \bigg], \nonumber \\  \pdv{\bm{k}}{t} &= \dfrac{1}{\hbar\mathcal{D}}\bigg[ \overleftrightarrow{\mathbf{R}} \bm{\nabla}_k \widetilde{\mathcal{E}} - \overleftrightarrow{\mathbf{S}}^T\bm{\nabla}_r \widetilde{\mathcal{E}}\bigg]
    \label{xidot}
\end{align}

\noindent
where to lowest order in spatial gradients $\mathbf{S}_{ij} \approx \delta_{ij}$, $\mathcal{D}\approx 1$, $\tilde{\mathcal{E}}\approx \mathcal{E}$,

\begin{equation}
    \mathbf{K}_{ij} \approx  \sum_{i,j=1}^3\epsilon_{ijk} \Omega_k^{K,\text{sc}} \,\,,\,\, \mathbf{R}_{ij} \approx  \sum_{i,j=1}^3\epsilon_{ijk} \Omega_k^{R,\text{sc}}
\end{equation}

\noindent
with $\bm{\Omega}^{K,\text{sc}}(\bm{r},\bm{k})=i\bm{\nabla}_k\times \bra{W_n(\bm{r},\bm{k})}\bm{\nabla}_k\ket{W_n(\bm{r},\bm{k})}$ and $\bm{\Omega}^{R,\text{sc}}(\bm{r},\bm{k})=i\bm{\nabla}_r\times \bra{W_n(\bm{r},\bm{k})}\bm{\nabla}_r\ket{W_n(\bm{r},\bm{k})}$ the semiclassical momentum space and position space Berry curvatures \cite{addison2025anomalous}.  For convenience we will suppress the band label $n$ on the quantities above.

To make connection with the position space Chern number, consider a system with momentum space translation symmetries for which $t^K_{\alpha\beta}(\bm{k}_i,\bm{k}_j)=t^K_{\alpha\beta}(\bm{k}_i-\bm{k}_j)$.  In this limit eigenfunctions of the Hamiltonian are localized modes $\ket{w(\bm{r}_i)}$.  In the thermodynamic limit equation \eqref{CRW} can be written as

\begin{equation}
    C_R=\dfrac{1}{2\pi}\sum_n\int d^2r \,f_n(\bm{r}) \bm{\Omega}^n_R(\bm{r})\cdot\bm{\hat{z}}
\end{equation}

\noindent
where $\bm{\Omega}^n_R(\bm{r})=i\bm{\nabla}_r\times \bra{w_n(\bm{r})}\bm{\nabla}_r\ket{w_n(\bm{r})}$ is the fully quantum mechanical position space Berry curvature in {\it band} $n$.  The curvature describes the evolution of the $U(1)$ phase of eigenstates $\ket{w_n(\bm{r})}$ across the system.  For these completely local systems $\ket{W_n(\bm{r},\bm{k})}\rightarrow \ket{w_n(\bm{r})}$ and $\bm{\Omega}^{R,\text{sc}}(\bm{r},\bm{k})\rightarrow \bm{\Omega}^n_R(\bm{r})$.

The semiclassical equations describe the relationship between the force on a wavepacket, $\bm{F}^{\text{sc}}=\hbar\partial_t\bm{k}$, the semiclassical position space Berry curvature, $\bm{\Omega}^{R,\text{sc}}(\bm{r},\bm{k})$, and the semiclassical velocity, $\bm{v}^{\text{sc}}(\bm{r},\bm{k})=\bm{\nabla}_k\tilde{\mathcal{E}}/\hbar$ via $\bm{F}^{\text{sc}}\sim\bm{v}^{\text{sc}}\times \bm{\Omega}^{R,\text{sc}}$.  At the fully quantum mechanical level this relationship is reflected in $\bm{j}^p\sim \bm{V}\times\bm{\chi}_H$, with $\bm{j}^p$ playing the role of the force density, $\bm{\chi}_H$ the average position space curvature  across the system or position space Chern number, and $\bm{V}$ the flow of particles (note that in 2D $\bm{\chi}_H=\chi_H\bm{\hat{z}}$ has a single nonzero component).

Most systems of course are not completely local $t^K_{\alpha\beta}(\bm{k}_i,\bm{k}_j)\neq t^K_{\alpha\beta}(\bm{k}_i-\bm{k}_j)$ and the notion of a local position space Berry curvature becomes ill-defined.  However, in the case of systems with a local long wavelength potential that breaks the position space translation symmetries, the semiclassical theory provides an approximate solution through $\bm{\Omega}^{R,\text{sc}}(\bm{r},\bm{k})$, and controlled by the small parameter $a/L_{s}$, where $L_s$ is the characteristic length scale of the local potential and $a$ the lattice constant.  The approximate semiclassical solution approaches the exact solution as $a/L_s\rightarrow 0$.

To see the connection between the fully quantum mechanical theory and the semiclassical theory, we consider a bilayer system where a momentum space Chern insulator is proximity coupled to the local magnetic moments of a skyrmion magnetic crystal.  We consider this 2D system on periodic boundary conditions with a Hamiltonian in the form of $\widehat{H}=\widehat{H}_R+\widehat{H}_K$ where

\begin{equation}
\widehat{H}_K=t_K\sum_{i=1}^N\sum_{\alpha,\beta=\uparrow,\downarrow}\widehat{c}^\dagger_{\alpha}(\bm{k}_i)(\bm{d}(\bm{k}_i)\cdot\bm{\sigma}_{\alpha\beta})\widehat{c}^{\phantom{\dagger}}_\beta(\bm{k}_i)
\end{equation}

\noindent
with $\bm{d}(\bm{k})=\sin(k_{x_i}a)\bm{\hat{x}}+\sin(k_{y_i}a)\bm{\hat{y}}+(\cos(k_{x_i}a)+\cos(k_{y_i}a)-\eta_K)\bm{\hat{z}}$, with $\bm{k}_i=k_{x_i}\bm{\hat{x}}+k_{y_i}\bm{\hat{y}}$ and $\widehat{H}_R$ is defined in equation \eqref{HSK}.

\begin{figure}
 \begin{centering}
\includegraphics[width=.48\textwidth]{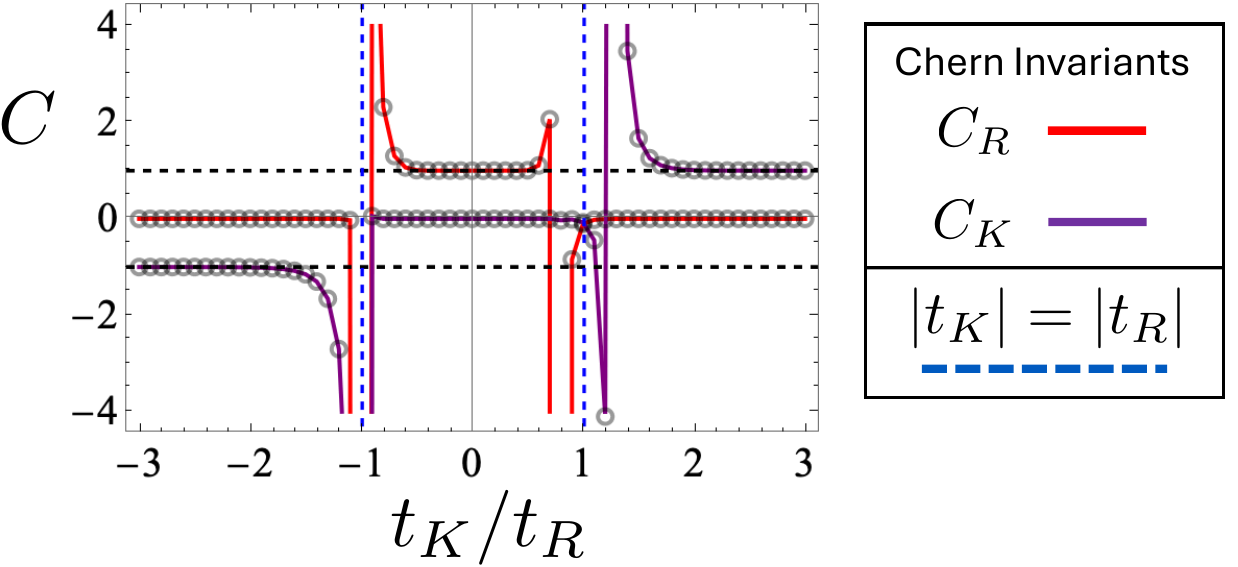}
\caption{{\bf Phase Space Chern Invariants:}  Position and momentum space Chern invariants as a function of the ratio of coupling strengths $t_K/t_R$ for a system of size $L=30a$ and $\eta_K=\eta_R=1$.  For $|t_K|\ll |t_R|$ the system is in a Chern insulating phase with $C_K=0$ and $C_R=1$.  As the coupling strengths become comparable $|t_K|\sim |t_R|$ (dashed blue lines) the system becomes gapless.  At large $|t_K| \gg |t_R|$ system is in a Chern insulating phase with $C_K=\text{sgn}(t_K/t_R)$.}
 \label{EComp}
 \end{centering}
\end{figure}

Here we investigate the topological properties of these insulators as a function of the coupling $t_R$, $t_K$, $\eta_R$, and $\eta_K$.  First we consider the momentum and position space Chern numbers as a function of $t_K/t_R$.  The Chern numbers are topological invariants whose value can change when the systems energy gap closes.  Figure \ref{EComp} plots $C_R$ and $C_K$ as a function $t_K/t_R$ for a system of size $L=30a$ with $\eta_R=\eta_K=1$.  We see for small $t_K$ the position space texture dominates the topology of the system exhibiting a phase with $C_R=1$ and $C_K=0$.  A phase transition occurs when the coupling to the position space and momentum space textures are of equal magnitude $t_R\sim t_K$ at which point the system's energy gap closes.  On the other side of the transition $t_K$ is the dominant energy scale and the topology of system is categorized by $C_K=\pm1$ and $C_R=0$.

\begin{figure}
 \begin{centering}
\includegraphics[width=.48\textwidth]{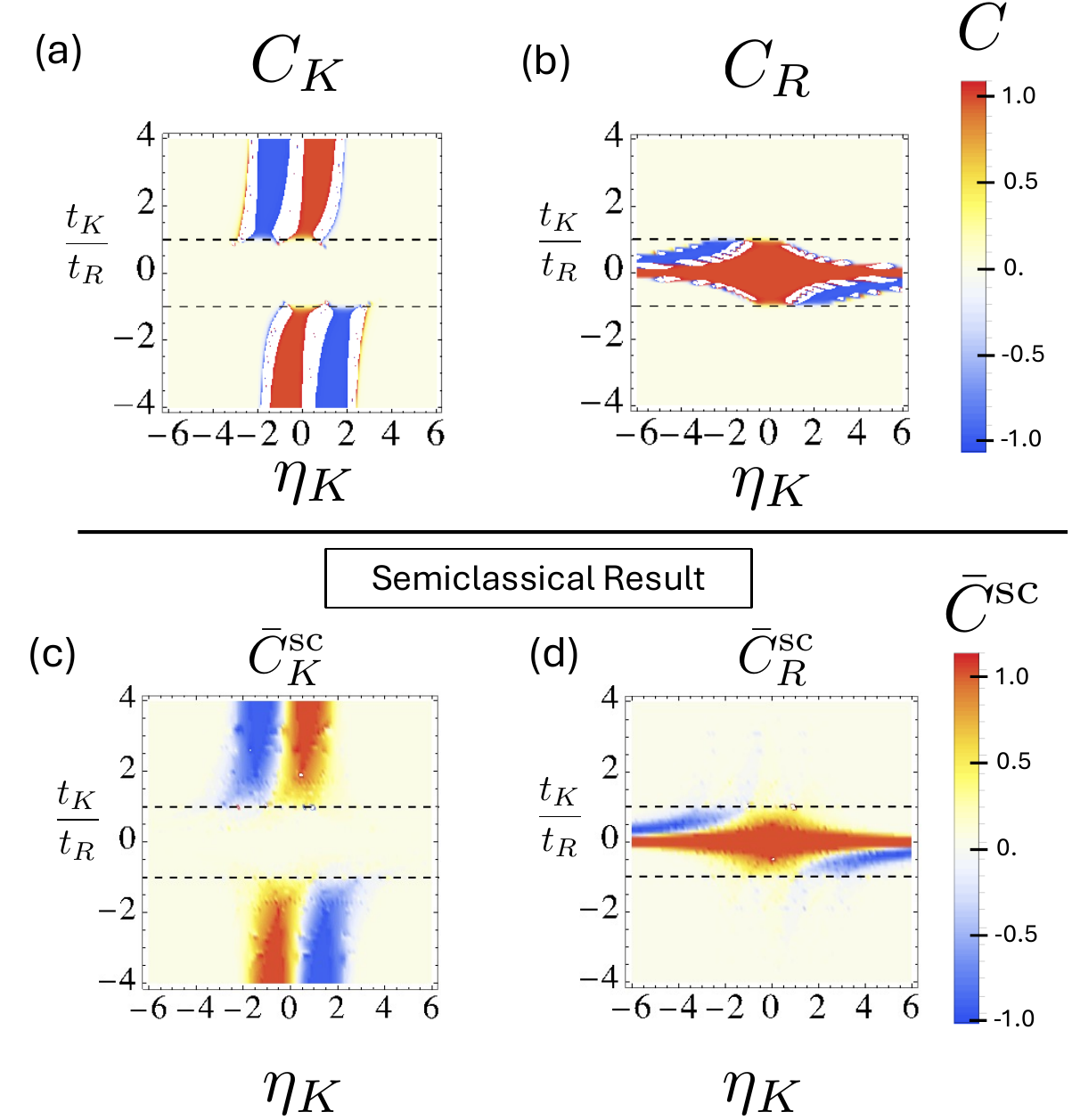}
\caption{{\bf Topological Phase Spaces and Semiclassical Results:}  Position and momentum space Chern invariants as a function of the ratio of coupling strengths $t_K/t_R$ and $\eta_K$ with $\eta_R=1$ for $L=20a$. (a)-(b) Full quantum mechanical calculation showing transition between topological phases.  Dashed lines show boundary where $t_R\sim t_K$ around which the dominant coupling strength ($t_K$ or $t_R$) changes.  Note that there is no region for which $C_K$ and $C_R$ are non-zero at the same time, indicating evidence for a topological obstruction preventing this type of state. (c)-(d) Semiclassical calculation of average Chern numbers across the system reproducing the diagram of the full quantum calculation in (a)-(b) with transitions between different topological sectors smoothed.  Here $L_{s}\sim L\gg a$ such that gradient corrections to the curvatures $\bm{\Omega}^n_K(\bm{\xi})$ and $\bm{\Omega}_{R}^n(\bm{\xi})$ are small.}
 \label{PS1}
 \end{centering}
\end{figure}

Looking toward the phase space parameterized by $\eta_K$ and $\eta_R$, figure \ref{PS1}(a)-(b) shows $C_K$ and $C_R$ as a function of $\eta_K$ and $t_K/t_R$ for $\eta_R=1$ and a system with size $L=20a$.  We see that there exists no region for which $C_K$ and $C_R$ are both non-zero suggesting there may be an obstruction for having a system where both invariants are non-zero.

The semiclassical Hamiltonian for this system is described by $t_{\alpha\beta}^{\text{sc}}(\bm{r},\bm{k})=(t_K\bm{d}(\bm{k})+t_R\bm{m}(\bm{r}))\cdot \bm{\sigma}_{\alpha\beta}$, with eigenvalues $\mathcal{E}_{\pm}(\bm{r},\bm{k})=\pm|t_K\bm{d}(\bm{k})+t_R\bm{d}(\bm{k})|$.  To make connection with the fully quantum theory we can define average semiclassical Chern numbers

\begin{align}
    \bar{C}^{\text{sc}}_K&=\dfrac{1}{2\pi V}\int d^2r\int d^2k \, \bm{\Omega}_-^{K,\text{sc}}(\bm{r},\bm{k})\cdot \bm{\hat{z}} \\
    \bar{C}^{\text{sc}}_R&=\dfrac{1}{2\pi V_{BZ}}\int d^2k\int d^2r \, \bm{\Omega}_-^{R,\text{sc}}(\bm{r},\bm{k})\cdot \bm{\hat{z}}
\end{align}

\noindent
where $V_{BZ}$ is the volume of the Brillouin zone.  Figure \ref{PS1}(c)-(d) shows the values of the semiclassical Chern numbers $\bar{C}^{\text{sc}}_K$ and $\bar{C}^{\text{sc}}_R$ as a function of $\eta_K$ and the ratio $t_K/t_R$ for $\eta_R=1$.  The semiclassical result is in good agreement with the full quantum calculation as $a/L_{sk}\sim 0.05\ll 1$ for this system leading to a good semiclassical approximation.  Moreover the integral of $\bm{\Omega}^{K,\text{sc}}(\bm{r},\bm{k})$ over the Brillouin zone or $\bm{\Omega}^{R,\text{sc}}(\bm{r},\bm{k})$ over the volume of the system is itself a topological invariant for insulating systems, and therefore insensitive to perturbations that do not close and reopen the energy gap.  If the quantum corrections to the semiclassical theory do not lead to a gap closure and reopening, we expect that this average semicalssical Chern number should well approximate the quantum result.

\section{Conclusion}

The position space Chern number, $C_R$, demonstrates the existence of topological invariants that challenge the canonical framework where momentum is the preferred 
basis in which to understand the topology of a system.  If we demand the system maintain other spacetime symmetries, by looking toward the position basis, new invariants analogous to, for example, the $Z_2$ Kane-Mele invariant, the $Z_2$ topological 1D inversion symmetric or 3D time-reversal invariant polarizations could be found \cite{kane2005z,fu2007topological,qi2008topological,hughes2011inversion}.

For the position space Chern number, systems hosting skyrmion phases are prime candidates for observing the edge states and bulk topological responses associated with nonzero $C_R$.  However, the Chern number is not limited to these systems, and classes of systems for which the internal degrees of freedom in a unit cell are larger than two, with no vector field representation, or associated winding number $N_{\text{sk}}$, also can serve as a platforms for finding systems with nonzero $C_R$.  In the case of momentum space Chern insulators a common theoretical and experimental tool for finding or engineering nonzero $C_K$ is looking for topological symmetry protected degeneracies in the band structure of a material and finding perturbations that break these symmetries leading to an insulating energy gap with topological properties \cite{wieder2016double,armitage2018weyl}.  In principle a similar technique could be applied to highly localized systems with topological crossings in the approximate position space spectrum of the material, $E_n(\bm{r})$.

To observe the topological edge states associated with $C_R\neq0$ requires the application of a perturbing momentum space wall.  Here we have provided a class of walls that lead to local position space coupling between sites, however it is unclear what the most experimentally viable potential barrier might be.  Beyond the electronic materials described here, observing these edgestates in the context of electromagnetic waveguides or mechanical devices may serve as easier platforms to design and implement these walls \cite{lu2014topological,kane2014topological,khanikaev2017two,ozawa2019topological,shah2024colloquium}.

In the presence of interactions that conserve particle number, $C_R$ remains well defined as long as the spectrum remains non-degenerate as equation \eqref{CRW} makes no reference to the types of terms in $\widehat{H}$.  In the context of degenerate systems $E_n=E_m$ for some $n,m$ its an open question of whether there is any type of interactions that can drive $\chi_H$ to fractionalize in ways similar to $\sigma_H$ in a fractional quantum Hall state \cite{lopez1991fractional,stormer1999fractional,hansson2017quantum}.  Determining the spectral flow within the degenerate manifold of the system's eigenspectrum as a function of the momentum space fluxes generated by nonzero $\bm{\mathcal{A}}^K$, could be a powerful tool for finding states where $\chi_H$ takes fractional values \cite{prodan2010entanglement}. 

Lastly, we note that the models discussed above seems to exhibit an inherent topological obstruction between the momentum space and position space Chern numbers as figure \ref{PS1}(a)-(b) exhibits no phase space region where $C_K$ and $C_R$ are both nonzero.  Clearly if the internal vector space  hosting the distinct topologies were uncoupled this apparent obstruction would vanish.  Looking toward the edge state picture for $C_R\neq 0$ we see, for example, in figure \ref{gausswall}(b), that the momentum space {\it edge} states possibly provide a potential pathway for allowing the canonical position space edge states to hybridize destroying their topological protection and ultimately leading to  $C_K=0$.  Interestingly, a semiclassical model of the form $t_{\alpha\beta}^{\text{sc}}(\bm{r},\bm{k})=\bm{d}(2\pi \bm{r}/aL+\bm{k})\cdot \bm{\sigma}_{\alpha\beta}$ has $ \bar{C}_K^{\text{sc}}=\bar{C}_R^{\text{sc}}\neq 0$ for a large range of $\eta_K$.  Whether or not this system provides a faithful representation at the fully quantum level is left for future work.  An open question, then, is if an obstruction exists how to best classify it, and if no obstruction exists what systems host a $C_K\neq0$ and $C_R\neq0$ phase and what the physical implications of such a state would be.

\section*{Acknowledgment}

The author thanks Xu Yang and Barry Bradlyn for insightful discussions related to this work.

\bibliography{Bibliography}

\appendix

\section{Alternative Definitions of Currents Through Momentum Space} \label{JPA}

Here we demonstrate the difference between the currents described in section \ref{PSCN} and alternative definitions for current flows in momentum space.  For simplicity consider a general noninteracting Hamiltonian that we may write as 

\begin{equation}
    \widehat{H}=\int d^3p d^3p' \,\dfrac{1}{2}\bigg(\widehat{c}^\dagger(\bm{p})t(\bm{p},\bm{p}')\widehat{c}(\bm{p'})+ \widehat{c}^\dagger(\bm{p}')t(\bm{p}',\bm{p})\widehat{c}(\bm{p})\bigg)
\end{equation}

\noindent
$\widehat{H}$ is invariant under the global gauge transformation $\widehat{c}^\dagger(\bm{p})\rightarrow \widehat{c}^\dagger(\bm{p})e^{i\theta}$.  Under a local gauge transformation $\theta\rightarrow \theta(\bm{p})$, to first order $\bm{\nabla}_p\theta(\bm{p})$ the change in the Hamiltonian $\Delta H$ is

\begin{align}
    \Delta\widehat{H}&=\int d^3p\bigg[\theta(\bm{p})\bm{\nabla}_p\bigg(\int d^3p' \dfrac{i}{2}(\bm{p}'-\bm{p}) \nonumber \\
&\bigg(\widehat{c}^\dagger(\bm{p})t(\bm{p},\bm{p}')\widehat{c}(\bm{p}') -\widehat{c}^\dagger(\bm{p}')t(\bm{p}',\bm{p})\widehat{c}(\bm{p})\bigg)\bigg)\bigg]
\end{align}

\noindent
where we have neglected boundary terms in the integrals.  For $\theta$ constant $\Delta \widehat{H}=0$, therefore we may define a local conserved current operator

\begin{align}
    \widehat{\bm{j}}^p(\bm{p})&= \int d^3p' \,\dfrac{i}{2}(\bm{p}'-\bm{p})\bigg(\widehat{c}^\dagger(\bm{p})t(\bm{p},\bm{p}')\widehat{c}(\bm{p}') \nonumber \\
    &-\widehat{c}^\dagger(\bm{p}')t(\bm{p}',\bm{p})\widehat{c}(\bm{p})\bigg)
\end{align}

\noindent
We note previous work has defined a momentum current operator as $\widehat{\bm{j}}(\bm{p})\sim \widehat{c}^\dagger(\bm{p})\widehat{\bm{p}} \widehat{c}(\bm{p})$, \cite{nalewajski2015probability,nalewajski2015quantum}.  Unlike the current $\bm{j}^p(\bm{p})=\text{Tr}(\widehat{\rho}(t) \widehat{\bm{j}}^p(\bm{p}))$, the currents stemming from the operator $\widehat{\bm{j}}(\bm{p})\sim \widehat{c}^\dagger(\bm{p})\widehat{\bm{p}} \widehat{c}(\bm{p})$ do not satisfy a local continuity equations and are not related to the global gauge invariance of $\widehat{H}$ described above.

The time component of the conserved current can be determined by coupling the system to a time dependent gauge transformation on the heisenburg field operators: $\widehat{\psi}^\dagger(\bm{r},t)=\widehat{\psi}^\dagger(\bm{r},t)e^{i\theta(t)}$.  For electronic systems the terms in the action that are modified take the form

\begin{equation}
    S\sim \int d^3rdt \, \widehat{\psi}^\dagger(\bm{r},t)i\partial_t\widehat{\psi}(\bm{r},t)
\end{equation}

\noindent
Again looking at first order changes with respect to $\theta(t)$ and neglecting boundary terms we have

\begin{align}
    \Delta S&\sim\int d^3rdt \,\theta(t) \partial_t ( \widehat{\psi}^\dagger(\bm{r},t)\widehat{\psi}(\bm{r},t)) \nonumber\\
    &=\int d^3pdt \,\theta(t) \partial_t ( \widehat{\psi}^\dagger(\bm{p},t)\widehat{\psi}(\bm{p},t))
\end{align}

\noindent
This leads to the expressions $n_r(\bm{r},t)=\bra{\bm{r}}\widehat{\rho}(t)\ket{\bm{r}}$ and $n_p(\bm{p},t)=\bra{\bm{p}}\widehat{\rho}(t)\ket{\bm{p}}$.

This procedure for defining a conserved current may be generalized to arbitrary interactions where $\widehat{H}$ need no longer be only quadratic in annihilation and creation operators and the interactions do not involve time derivatives of the field operators.  If terms in the Hamiltonian have an unequal amount of annihilation and creation operators the system does not conserve the number of quanta and there is no longer a global gauge symmetry of this form or associated conserved current.

In classical Hamiltonian mechanics, the probability distribution $\rho(\bm{r},\bm{p},t)$ of a particle satisfies a Louisville equation

\begin{equation}
    \partial_t \rho +\bm{\nabla}_r\cdot\bm{j}_r +\bm{\nabla}_p\cdot\bm{j}_p =0
\end{equation}

\noindent
A similar equation generalizing flows in a quantum mechanical phase space can be found utilizing the Wigner transform and Moyal product of the von Neumann equation for the density matrix.  The resultant quantum Louisville equation allow for identification of the quantum versions of $\bm{j}_r$ and $\bm{j}_p$ and are not derived from a gauge symmetry of the system as the currents defined above in the main text \cite{wigner1932quantum,skodje1989flux,ballentine1994inadequacy,steuernagel2013wigner,kakofengitis2017wigner,valtierra2020quasiprobability}.

\end{document}